\documentclass[aps,12pt,prl,twocolumn,]{revtex}
\usepackage{graphicx}
\title{
Thermalization temperature in Pb+Pb
collisions at SpS energy from hadron
yields and midrapidity $p_t$
distributions of hadrons and direct
photons}

\author{D. Yu. Peressounko\cite{address}
        and Yu. E. Pokrovsky}

\address{
Russian Research Center "Kurchatov
Institute", Kurchatov sq. 1, 123182
Moscow, Russia.}

\begin{document}

\maketitle

\begin{abstract}
In the frame of 2+1 Bjorken hydrodynamics we describe
simultaneously hadron yields and midrapidity $p_t$
distributions of direct photons, unflavored, and strange
hadrons measured in Pb+Pb collisions at SpS. We find, that a
reasonably quantitative description of these data is
possible only if we introduce some radial velocity at the
beginning of the one-fluid hydrodynamic stage. We fix
uniquely all parameters of this model, and estimate an
initial temperature of the just thermalized state
$T_{in}=200{+40\atop-20}\,MeV$ in the case of EoS with phase
transition to Quark-Gluon Plasma, and
$T_{in}=230{+60\atop-30}\, MeV$ in the pure hadronic
scenario. We conclude that SpS data considered in our study
can not distinguish between a production of QGP, and its
absence.
\end{abstract}

\vspace{0.3cm}



A finite volume of rapidly expanding Quark-Gluon Plasma
(QGP) can be possibly created for a very short time in
ultrarelativistic heavy ion collisions as an almost locally
equilibrated thermodynamic state of deconfined quarks and
gluons with strong residual interaction. Taking into account
a very short time available for QGP formation in this
process a necessary condition of its creations is that an
initial (thermalization) temperature $T_{in}$ have to be
visibly higher than the quark-hadron phase transition
temperature, which is estimated from lattice calculations to
be $T_c=150-170\,MeV$ \cite{lattice}. Therefore an
estimation of a temperature at which the local thermal
equilibrium is just established -- thermalization
temperature -- is of great importance.

There are a lot of attempts to extract this temperature from
direct photon momentum distributions measured in S+Au
\cite{S+Au->spectrum} and Pb+Pb \cite{Pb+Pb->spectrum}
collisions. However, to extract the thermalization
temperature from experimental data reasonably well, it is
necessary to separate accurately the collective and chaotic
motions. To do this one needs to know the evolution of the
collision starting from the first interactions till the
final particle stage. Unfortunately, the most part of the estimations
of the thermalization temperature \cite{S+Au->Tin,Pb+Pb->Tin} 
are based on hydrodynamic description with too simplified initial 
conditions (zero initial radial collective velocity), and other 
model parameters fixed partially by some experimental data, and
partially -- ad hoc from more or less reasonable considerations. 
Only in the recent paper \cite{Pb+Pb->TinVin} a first attempt 
to estimate the influence of nonzero initial radial velocity 
was undertaken.

In contrast to this, in the present letter we use initial
conditions for hydrodynamic model in the most general form,
describe simultaneously a large amount of available
experimental data, measured in Pb+Pb collisions at SpS
(hadron yields, midrapidity $p_t$ distributions of direct
photons, unflavored and strange hadrons), fix uniquely all
model parameters from these data and obtain reliable
estimate of the thermalization temperature with its
uncertainty concerned with experimental data bars.

In this way we find, that if we use a realistic equation of
state (EoS) of hadronic gas incorporating contributions of
all known hadrons and resonances, then to describe
midrapidity $p_t$ distributions of final hadrons it is
necessary to assume an essentially nonzero initial radial
collective velocity at the beginning of the one-fluid
hydrodynamic stage. This velocity is created during the
pre-equilibrium (two-fluid) stage of the collision due to
significant radial energy density gradient coming from the
shapes of the colliding nuclei. Nevertheless, in all
previous estimations of the thermalization temperature this
initial radial velocity was absolutely neglected, and only
the simplest case (zero initial velocity) was considered.

Evaluating the direct photon yield we assume the following
scenario of the collision. Initially, on the pre-equi\-li\-bri\-um
stage, colliding nuclei penetrate through each other like
two fluids with a friction. On this stage the `prompt'
photons are emitted and radial collective velocity is
created. If the thermalization time is not negligibly small then 
the both these effects should be taken into account. 
The next stage is a 
one-fluid-like expansion of the locally thermalized matter 
with an emission of the `thermal' photons and its decay 
onto final hadrons.

To begin with we calculate contribution of the prompt
photons. We take
into account Compton, annihilation and
bremsstrahlung processes. We use GRV-94 structure functions
\cite{GRV-94}, two-loop expression for $\alpha_s$ and
K-factor, K=2, accounting higher order corrections.
Comparing calculated yield of prompt photons in pp and pA
collisions with experimental data at $\sqrt{s}= 19$ GeV --
the closest available c.m. energy to Pb+Pb at SpS
($\sqrt{s}=17.3$ GeV) we find, that both experimental data
and theoretical predictions have approximately the same
slope, but difference in normalization reaches factor
$\approx 7$. Nowadays this difference is attributed to the
intrinsic transverse momentum of colliding partons
\cite{prompt-transverce}. In this letter we account these
corrections by introducing a proper factor, assuming that it
is independent on $\sqrt{s}$ while moving from pA to Pb+Pb
energy. To go from pA to AA collisions we need the number of
nucleon-nucleon collisions occurred during the
interpenetration of the colliding nuclei. In our
calculations we use Glauber approximation, what gives in the
case of Pb+Pb collision $N_{Gl}\approx 3\cdot A$. Comparison
with WA98 data \cite{Pb+Pb->spectrum} shows, that prompt
photons contribute approximately 30\% of total direct photon
yield.

To evaluate yield of thermal photons we integrate average
emission rate of direct photons from unit volume per unit
time over a space-time evolution of hot matter calculated in
a one-fluid hydrodynamics. In the present analysis we use
recently calculated emission rates accounting next to
leading effects in QGP \cite{Aurenche} and dominating
contribution in hadronic gas
-- reaction $\pi\rho\to a_1\to \pi\gamma$ \cite{Shuryak}.
For description of the evolution we use 2+1 Bjorken
hydrodynamics, which works sufficiently well at midrapidity.
It has several parameters which should be fixed to define
the evolution: a) initial conditions: initial temperature
$T_{in}$, initial time $\tau_{in}$, initial energy density
and velocity distributions; b) EoS of hot matter: transition
temperature $T_c$ (if EoS includes phase transition),
chemical freeze-out temperature $T_{ch}$ and thermal
freeze-out temperature $T_f$.

For the shape of energy density distribution we use
Woods-Saxon distribution, integrated along beam axis. As for
initial distribution of radial velocity, it deserve separate
discussion. As we find, to describe momentum distributions
of final hadrons using realistic EoS it is necessary to
introduce some radial velocity already at the beginning of
the one-fluid hydrodynamics. This velocity is created during
interpenetration of colliding nuclei due to significant
radial energy density gradient, originated from spherical
shapes of colliding nuclei. Two-fluid hydrodynamics or
kinetic models might predict value and radial dependence of
this collective velocity but, in this letter we use
`phemenologic' approach and use a simple linear distribution
$v(r)=(r/R)\cdot \Theta(R-r) \cdot V^{max}$ characterized by
one new parameter -- $V^{max}$ -- the velocity on the
surface of the just thermalized system. We find, that our
results are not sensitive to reasonable variations of the
shape of this distribution.

Concerning equation of state we consider two different
cases: EoS with phase transition to QGP, and pure hadronic
EoS. In the case of phase transition for QGP phase we use
EoS of ideal gas of massless quarks and gluons with
degeneracy $g=41.5$, what corresponds to QGP consisting of
2.5 massless quark flavors and gluons. To calculate EoS of
hadronic phase both in the case of phase transition and
pure hadronic scenario we take into account contributions of
all hadrons listed in the Particle Data Book \cite{PDB}.
Inclusion of heavy hadrons leads to significant reduction of
speed of sound in the hadronic phase: $v_s^2\approx 0.15$ in
our case should be compared with $v_s^2\approx 0.33$ for
massless hadronic gas or $v_s^2\approx 0.25$ for pure pion
gas. The decreasing of speed of sound results in decreasing
of acceleration in hadronic gas and stepper slopes of $p_t$ 
distributions of final hadrons \cite{hydro-reach-EoS}.

To model freeze-out of hadronic gas we use two different
parameters: chemical ($T_{ch}$) and thermal ($T_f$) 
freeze-out temperatures. It is known, that
due to large difference in cross-sections of elastic and
inelastic reactions in hadronic gas, the chemical 
freeze-out takes place before
the thermal one. Following paper \cite{Munzinger} we use the
following values of the temperature of chemical freeze-out,
and baryon, strange and $I_3$ chemical potentials --
$T_{ch}=168\, MeV$, $\mu_b=266 \, MeV$, $\mu_s=71\, MeV$ and
$\mu_3=-5\, MeV$. As a result we well describe yields of
$\pi$, $K$, $\eta$, $p$, $\Lambda$, $\Xi$ and $\Omega$
hadrons measured by NA44 \cite{NA44-hadr}, NA49
\cite{NA49-hadr}, WA97 \cite{WA97-hadr} and WA98
\cite{WA98-pi0} collaborations in Pb+Pb collisions at SpS
energy.

A description of our results we start from pure hadronic
scenario. Despite the belief that hadronic gas can not exist
at temperatures higher than $\sim 200$ MeV (because of the
essential overlapping of hadrons in it) we adopt its
existence and consider the consequences of this assumption
from the point of view of description of direct photon
production.
To begin with, we assume zero initial radial velocity and
try to describe momentum distributions of final hadrons. We
choose some initial temperature and calculate initial time
to describe multiplicity of final $\pi^0$. After that we
perform the hydrodynamic calculations with various values of
chemical and thermal freeze-out temperatures and obtain
$\chi^2_\pi$ of the fit of the calculated to the
experimental $\pi^0$ distributions as a function of $T_{ch}$
and $T_f$. We find, that it is
impossible to describe midrapidity $p_t$ distributions of
$\pi^0$ within any reasonable set of model parameters:
because of the much steeper slope of the calculated $p_t$
distribution of $\pi^0$ with respect to the experimental
one, we obtain $\chi^2_\pi\sim 10^2/point$. We find the same
situation for all initial temperatures in the range $160\,
MeV <T_{in}< 300\, MeV$. A reasonable way to improve this
situation is to introduce some initial radial velocity to
the one-fluid hydrodynamic stage.

\begin{figure}[htb]
\begin{center}
\includegraphics[width=80mm]{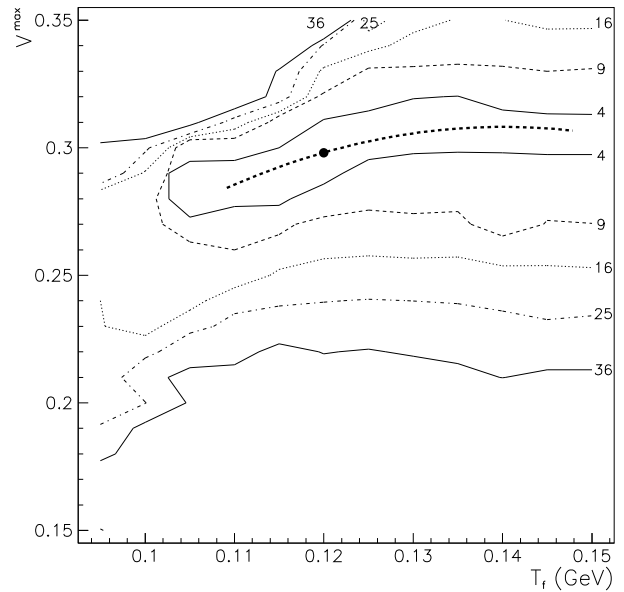}
\end{center}
\caption{
Levels of constant $\chi^2_\pi$ of the fit of the ratio of
calculated to measured $p_t$ distributions of $\pi^0$ as a
function of maximal initial velocity ($V^{max}$) and thermal
freeze-out temperatures ($T_f$) for pure hadronic gas EoS,
$T_{in}=230$ MeV.}
\label{fit-hvel}
\end{figure}

On the next step, we repeat calculations with nonzero
initial radial velocity. Again we choose initial
temperature, fix initial time from the multiplicity
considerations, fix temperature of chemical freeze-out
$T_{ch}=168\, MeV$ to describe hadron yields and plot levels
of constant $\chi^2_\pi$ in the $T_f-V^{max}$ coordinates -- see
fig.\ \ref{fit-hvel}. The function $\chi^2_\pi(T_f,V^{max})$ has
a shape of a valley with steep walls, while any point on its
bottom (thick line), corresponds to a good description of
midrapidity $p_t$ distributions of $\pi^0$ with $\chi^2_\pi\sim
1/point$. To fix a point on this line we use spectra of
heavier hadrons ($K$, $p$, $\Lambda$, $\Xi$). 
It was shown \cite{col-vel-hadr} that
spectra of the heavier hadrons can be described well if we
assume freeze-out temperature $T_f=120\, MeV$.

\begin{figure}[htb]
\begin{center}
\includegraphics[width=82mm]{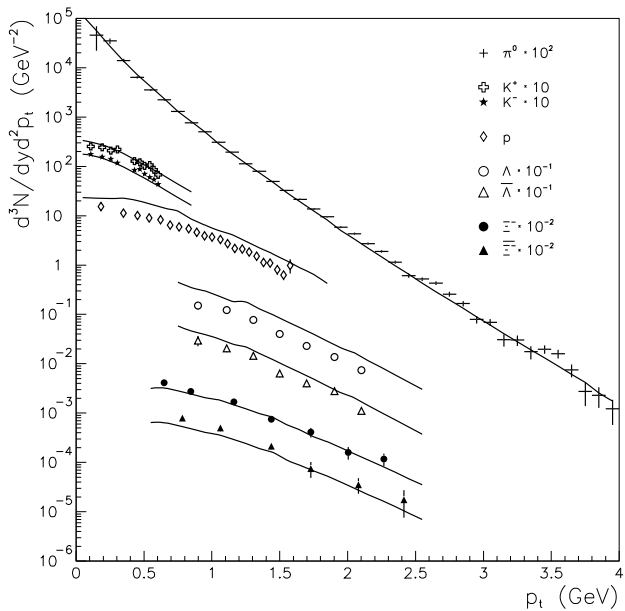}
\end{center}
\caption{
Comparison of experimental $p_t$
distributions of several hadrons with
hydrodynamic predictions for pure
hadronic gas EoS with $T_{in}=230\,MeV$
and initial radial velocity
$V^{max}=0.26$.}
\label{spectra}
\end{figure}

The description of the hadron $p_t$ distributions at
midrapidity obtained in this way is shown on the
fig.~\ref{spectra}. One can see, that shapes of all
available experimental spectra: $\pi^0$ \cite{WA98-pi0}, $K$
and $\Lambda$ \cite{K-Lambda-Spectra}, protons
\cite{p-Spectrum}, and $\Xi$ \cite{Xi-Spectrum} are
described very well, although we overestimate total yields
of $p$ and $\Lambda$ in this rough model.

\begin{figure}[htb]
\begin{center}
\includegraphics[width=82mm]{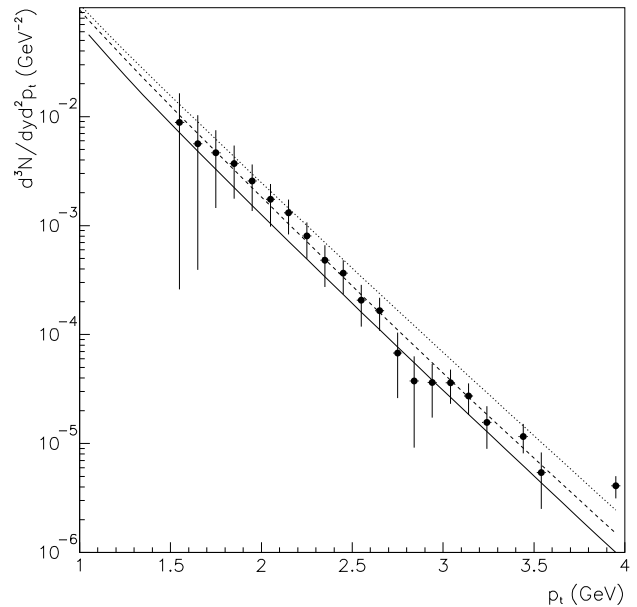}
\end{center}
\caption{
Yield of direct photons in Pb+Pb collisions at SpS
energy with pure hadronic gas EoS for different
$T_{in}$: 200 (solid line), 230 (dashed line) and
300 MeV (dotted line). Dots -- WA98 data.}
\label{dir-hvel}
\end{figure}

To fix finally all model parameters we estimate $T_{in}$
using $p_t$ distribution of direct photons. We depict on the
fig.~\ref{dir-hvel} the sum of the $p_t$ distribution of
prompt and thermal photons, calculated at three initial
temperatures. For each initial temperature we repeat
calculations, described above and obtain the following sets
of parameters:
\begin{center}
\begin{tabular}{cccl}
\hline
$T_{in}$, MeV ~& ~$\tau_{in}$, fm/c~~&~~~$V^{max}$, c~~~ & Curve on fig.~\ref{dir-hvel}\\       
\hline
 200     & 1.4          &  0.36     & ~~solid  \\
 230     & 0.6          &  0.29     & ~~dashed \\
 300     & 0.12         &  0.23     & ~~dotted \\
\hline
\end{tabular}
\end{center}

We find, an agreement with experimental data at initial
temperature $T_{in}=230{+60\atop-30}\,MeV$:
$T_{in}=230\,MeV$ corresponds to $\chi_{\pi}^2=3.7/point$,~
$\chi_{\gamma}^2=0.67/point$, while curves, calculated for
$T_{in}=200\, MeV$
($\chi_{\pi}^2=4.9/point$,~$\chi_{\gamma}^2=0.67/point$),
and $T_{in}=300\, MeV$ ($\chi_{\pi}^2=3.5/point$,
$\chi_{\gamma}^2=2.3/point$) are still within experimental
errors. The
reason of this weak sensitivity to the initial temperature
is following. As far as we introduce the initial radial
velocity, all time stages contribute with more or less the
same effective slopes. But, the difference between initial
time for $T_{in}=300$ MeV -- $\tau_{in}=0.12$ fm/c, and for
$T_{in}=230$ MeV -- $\tau_{in}=0.6$ fm/c is not large, and
therefore contribution of this stage is not very important
with respect to the subsequent evolution.

We confirm the result found in \cite{Pb+Pb->Tin} -- `reach' hadron
gas EoS allows to describe the experimental $p_t$
distribution of direct photons \cite{Pb+Pb->spectrum} within  pure
hadronic scenario. Our best fit value of $T_{in}=230{+60\atop-30}\,
MeV$ is smaller, than the initial temperature estimated in
\cite{Pb+Pb->Tin} $T_{in}\approx 260\, MeV$. There are two
reasons of this discrepancy: first is, the initial radial
velocity which we included to describe final hadron spectra.
In contrast to our approach, authors of \cite{Pb+Pb->Tin}
made not attempt to describe hadron spectra, and did not
consider a nonzero initial radial velocity. The second
reason is different normalization to final hadron
multiplicity. The authors of \cite{Pb+Pb->Tin} used the well
known Bjorken formula, relating initial time and temperature
with final multiplicity (formula (5) in \cite{Pb+Pb->Tin}).
But, this formula is valid only for massless final
particles and being applied to real hadronic gas leads to an
underestimation of the initial time and, as a consequence,
of the thermal photon yield.

Now let us consider EoS incorporating first
order phase transition. In this case there is an additional
parameter $T_c$ which value is estimated from lattice QCD
simulations to be $150-170~MeV$ \cite{lattice}. On the 
other hand $T_c\ge T_{ch}$. Therefore
we use $T_c=T_{ch}=168~MeV$ to consider an extreme case with
the most intensive QGP production.

As in the case of pure hadronic gas, first we try to
describe experimental $p_t$ distribution of $\pi^0$ with
zero initial radial velocity, and find almost the same
situation: for any reasonable set of model parameters the
calculated $p_t$ distribution of $\pi^0$ goes much steeper,
and result in unreasonably large $\chi^2_\pi\sim 10^2/point$. So,
we find again, that it is necessary to introduce initial
radial velocity to describe momentum distributions of final
hadrons. Repeating the same procedure as for pure hadronic
gas EoS we obtain the following sets of model parameters: 

\begin{center}
\begin{tabular}{cccl}
\hline
$T_{in}$, MeV ~~&~~$\tau_{in},$ fm/c~~ &~~~ $V^{max},$ c~~~& Curve on fig.~\ref{dir-vel}\\
\hline
 180     & 2.0        &  0.40     & ~~solid       \\
 200     & 1.5        &  0.38     & ~~dashed      \\
 230     & 1.0        &  0.29     & ~~dotted      \\
 250     & 0.7        &  0.26     & ~~dash-dotted \\
\hline
\end{tabular}
\end{center}

We find the best agreement with experimental data at $T_{in}=
200{+40\atop-20}$ MeV ($\chi_{\pi}^2=3.3/point$,~
$\chi_{\gamma}^2=0.8/point$), what means very short QGP
phase and much more prolonged mixed phase.

\begin{figure}[htb]
\begin{center}
\includegraphics[width=82mm]{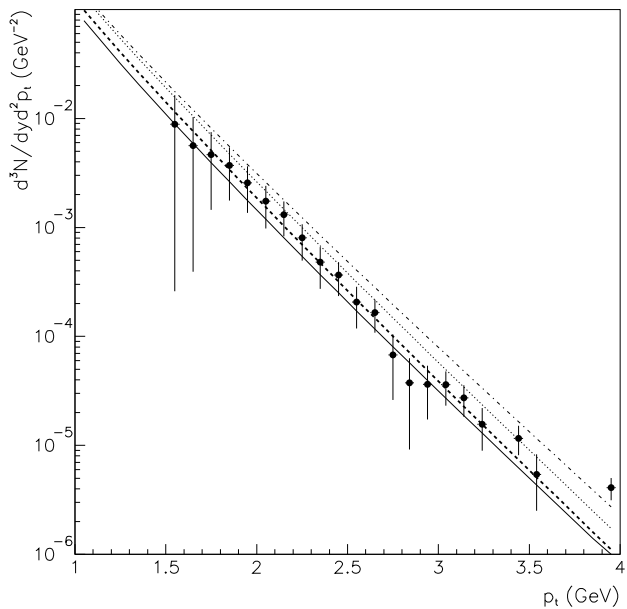}
\end{center}
\caption{
Direct photon yields calculated for EoS with phase
transition and linear initial velocity
distribution at $T_{in}=180$ MeV (solid line),
$T_{in}=200$ MeV (dashed line), $T_{in}=230$ MeV
(dotted line) and $T_{in}=250$ MeV (dash-dotted
line).}
\label{dir-vel}
\end{figure}

To conclude, we use 2+1 Bjorken hydrodynamics to analyze
hadron yields, and midrapidity $p_t$ distributions of
hadrons and direct photons, measured in Pb+Pb collision at
SpS energy. We find, that calculations with zero initial
velocity, and EoS of hadronic gas accounting contribution of
all known hadrons result in too soft $p_t$ distributions of
final hadrons, and a reasonable way to describe experimental
slopes is to introduce nonzero radial velocity at the
beginning of the one-fluid hydrodynamic stage. This initial
radial velocity, which might be produced during the pre-equilibrium
stage because of the strong radial energy density gradient
originated from spherical shapes of colliding nuclei, is
not small ($\approx0.3~c$ near the surface). Therefore the
pre-equilibrium stage is expected to be long enough, and, in
particular the `prompt photons' have to be taken into
account.

An introduction of the initial radial velocity to this model
allows to describe simultaneously direct photon and hadron
$p_t$ distributions for the both EoS including phase
transition and without it. We estimate an initial
temperature of the just thermalized state
$T_{in}=200{+40\atop-20}\, MeV$ in the case of EoS with
phase transition to QGP, and
$T_{in}=230{+60\atop-30}\, MeV$ in the pure hadronic
scenario. Because the lower bounds of the estimated
thermalization temperature are very close to the expected
transition temperature we conclude that the SpS data
considered in this letter can not distinguish between a
production of QGP, and its absence: The initial temperature
in central Pb+Pb collisions at SpS is too small, and
experimental errors for the direct photon $p_t$
distributions are too large.

This work was supported by the INTAS under Contract INTAS-97-0158, and by grant RFBR 00-15-96590.

\end{document}